# On the Approximation of the Sum of Lognormals by a Log Skew Normal Distribution

Marwane Ben Hcine[1] and Ridha Bouallegue[2]

[1,2]Innovation of Communicant and Cooperative Mobiles Laboratory, INNOV'COM
Sup'Com, Higher School of Communication
Univesity of Carthage
Ariana, Tunisia
[1]marwen.benhcine@supcom.tn
[2]ridha.bouallegue@supcom.rnu.tn

## ABSTRACT

*Several methods have been proposed to approximate the sum of lognormal RVs. However the accuracy of each method relies highly on the region of the resulting distribution being examined, and the individual lognormal parameters, i.e., mean and variance. There is no such method which can provide the needed accuracy for all cases. This paper propose a universal yet very simple approximation method for the sum of Lognormals based on log skew normal approximation. The main contribution on this work is to propose an analytical method for log skew normal parameters estimation. The proposed method provides highly accurate approximation to the sum of lognormal distributions over the whole range of dB spreads for any correlation coefficient. Simulation results show that our method outperforms all previously proposed methods and provides an accuracy within 0.01 dB for all cases.*

## 1. INTRODUCTION

Multipath with lognormal statistics is important in many areas of communication systems. With the emergence of new technologies (3G, *LTE*, *WiMAX*, Cognitive Radio), accurate interference computation becomes more and more crucial for outage probabilities prediction, interference mitigation techniques evaluation and frequency reuse scheme selection. For a given practical case, Signal-to-Interference-plus-Noise (SINR) Ratio prediction relies on the approximation of the sum of correlated lognormal RVs. Looking in the literature; several methods have been proposed in order to approximate the sum of correlated lognormal RVs. Since numerical methods require a time-consuming numerical integration, which is not adequate for practical cases, we consider only analytical approximation methods. Ref [1] gives an extension of the widely used iterative method known as Schwartz and Yeh (SY) method [2]. Some others resources uses an extended version of Fenton and Wilkinson methods [3-4]. These methods are based on the fact that the sum of dependent lognormal distribution can be approximated by another lognormal distribution. The non-validity of this assumption at distribution tails, as we will show later, is the main raison for its fail to provide a consistent approximation to the sum of correlated lognormal distributions over the whole range of dB spreads. Furthermore, the accuracy of each method depends highly on the region of the resulting distribution being examined. For example, Schwartz and Yeh (SY) based methods provide acceptable accuracy in low-precision region of the Cumulative Distribution Function (CDF) (i.e., 0.01–0.99) and the

Fenton–Wilkinson (FW) method offers high accuracy in the high-value region of the CDF (i.e., 0.9–0.9999). Both methods break down for high values of standard deviations. Ref [5] propose an alternative method based on Log Shifted Gamma (LSG) approximation to the sum of dependent lognormal RVs. LSG parameters estimation is based on moments computation using Schwartz and Yeh method. Although, LSG exhibits an acceptable accuracy, it does not provide good accuracy at the lower region.

In this paper, we propose a very highly accurate yet simple method to approximate the sum of lognormal RVs based on Log Skew Normal distribution (LSN). LSN approximation has been proposed in [6] as a highly accurate approximation method for the sum of independent lognormal distributions, Furthermore a modified LSN approximation method is proposed in [7]. However, LSN parameters estimation relies on a time-consuming Monte Carlo simulation and the proposed approach is limited to the independent case. The main contribution on this work is to provide a simple analytical method for LSN parameters estimation without the need for a time-consuming Monte Carlo simulation or curve fitting approximation and extend previous approaches to the correlated case. Our analytical fitting method is based on moments and tails slope matching for both distributions. This work can be seen as extension to the correlated case for our work done in [8].

The rest of the paper is organized as follows: In section 2, a brief description of the lognormal distribution and sum of correlated lognormal distributions is given. In section 3, we introduce the Log Skew Normal distribution and its parameters. The validity of lognormal assumption for the sum of lognormal RVs at distribution tails is discussed in section 4. In section 5, we use moments and tails slope matching method to estimate LSN distribution parameters. In section 6, we provide comparisons with well-known approximation methods (i.e. Schwartz and Yeh, Fenton–Wilkinson, LSG) based on simulation results. In section 7, we give an example for outage probability calculation in lognormal shadowing environment based on our method.

The conclusion remarks are given in Section 8.

## 2. SUM OF CORRELATED LOGNORMAL RVS

Given $X$, a Gaussian RV with mean $\mu_X$ and variance $\sigma_X^2$, then $L=e^X$ is a lognormal RV with a Probability Density Function (PDF):

$$f_L(l,\mu_X,\sigma_X) = \begin{cases} \frac{1}{\sqrt{2\pi}l\sigma_X}\exp(-\frac{1}{2\sigma_X^2}[\ln(l)-\mu_X]^2) & l>0 \\ 0 & \text{otherwise} \end{cases} \qquad (1)$$

Usually $X$ represents power variation measured in dB. Considering $X_{dB}$ with mean $\mu_{dB}$ and variance $\sigma_{dB}^2$, the corresponding lognormal RV $L = 10^{\frac{X_{dB}}{10}}$ has the following pdf:

$$f_L(l,\mu_{dB},\sigma_{dB}) = \begin{cases} \frac{1}{\sqrt{2\pi}l\sigma_{dB}}\exp(-\frac{1}{2\sigma_{dB}^2}[10\log(l)-\mu_{dB}]^2) & l>0 \\ 0 & \text{otherwise} \end{cases} \qquad (2)$$

where $\mu_{dB} = \frac{\mu_X}{\xi}$, $\sigma_{dB} = \frac{\sigma_X}{\xi}$

and $\xi = \frac{\ln(10)}{10}$

The first two central moments of $L$ may be written as:

$$m = e^{\mu} e^{\sigma^2/2}$$
$$D^2 = e^{2\mu} e^{\sigma^2}(e^{\sigma^2} - 1) \qquad (3)$$

Correlated Lognormals sum distribution corresponds to the sum of dependent lognormal RVs, i.e.

$$\Lambda = \sum_{i=1}^{N} L_n = \sum_{i=1}^{N} e^{X_n} \qquad (4)$$

We define $\vec{L} = (L_1, L_2 ... L_N)$ as a strictly positive random vector such that the vector $\vec{X} = (X_1, X_2 ... X_N)$ with $X_j = \log(L_j)$, $1 \leq j < N$ has an n-dimensional normal distribution with mean vector $\mu = (\mu_1, \mu_2 ... \mu_N)$ and covariance matrix $M$ with $M(i,j) = Cov(X_i, X_j)$, $1 \leq i < N, 1 \leq j < N$. $\vec{L}$ is called an n-dimensional log-normal vector with parameters $\vec{\mu}$ and $M$.
$Cov(L_i, L_j)$ may be expressed as [9, eq. 44.35]:

$$Cov(L_i, L_j) = e^{\mu_i + \mu_j + \frac{1}{2}(\sigma_i^2 + \sigma_j^2)} (e^{M(i,j)} - 1) \qquad (5)$$

The first two central moments of $\Lambda$ may be written as:

$$m = \sum_{i=1}^{N} m_i = \sum_{i=1}^{N} e^{\mu_i} e^{\sigma_i^2/2} \qquad (6)$$

$$D^2 = \sum_{i=1,j=1}^{N} Cov(L_i, L_j) = \sum_{i=1,j=1}^{N} e^{\mu_i + \mu_j + \frac{1}{2}(\sigma_i^2 + \sigma_j^2)} (e^{M(i,j)} - 1) \qquad (7)$$

## 3. LOG SKEW NORMAL DISTRIBUTION

The standard skew normal distribution was firstly introduced in [10] and was independently proposed and systematically investigated by Azzalini [11]. The random variable $X$ is said to have a scalar $SN(\lambda, \varepsilon, \omega)$ distribution if its density is given by:

$$f_X(x; \lambda, \varepsilon, \omega) = \frac{2}{\omega} \varphi(\frac{x - \varepsilon}{\omega}) \phi(\lambda \frac{x - \varepsilon}{\omega}) \qquad (8)$$

Where $\varphi(x) = \frac{e^{-x^2/2}}{\sqrt{2\pi}}, \quad \phi(x) = \int_{-\infty}^{x} \varphi(\zeta) d\zeta$

With $\lambda$ is the shape parameter which determines the skewness, $\varepsilon$ and $\omega$ represent the usual location and scale parameters and $\varphi$, $\phi$ denote, respectively, the pdf and the cdf of a standard Gaussian RV.
The CDF of the skew normal distribution can be easily derived as:

$$F_X(x; \lambda, \varepsilon, \omega) = \phi(\frac{x - \varepsilon}{\omega}) - 2T(\frac{x - \varepsilon}{\omega}, \lambda) \qquad (9)$$

Where function $T(x, \lambda)$ is Owen's $T$ function expressed as:

$$T(x, \lambda) = \frac{1}{2\pi} \int_0^\lambda \frac{\exp\left\{-\frac{1}{2}x^2(1+t^2)\right\}}{(1+t^2)} dt \tag{10}$$

A fast and accurate calculation of Owen's T function is provided in [12].

Similar to the relation between normal and lognormal distributions, given a skew normal RV $X$ then $L = 10^{\frac{X_{dB}}{10}}$ is a log skew normal distribution. The cdf and pdf of $L$ can be easily derived as:

$$f_L(l; \lambda, \varepsilon_{dB}, \omega_{db}) = \begin{cases} \frac{2}{\xi \omega_{db} l} \varphi(\frac{10\log(l) - \varepsilon_{dB}}{\omega_{db}}) \phi(\lambda \frac{10\log(l) - \varepsilon_{dB}}{\omega_{db}}) & l > 0 \\ 0 & \text{otherwise} \end{cases} \tag{11}$$

$$F_L(l; \lambda, \varepsilon_{dB}, \omega_{db}) = \begin{cases} \phi(\frac{10\log(l) - \varepsilon_{dB}}{\omega_{db}}) - 2T(\frac{10\log(l) - \varepsilon_{dB}}{\omega_{db}}, \lambda) & l > 0 \\ 0 & \text{otherwise} \end{cases} \tag{12}$$

Where $\varepsilon_{dB} = \frac{\varepsilon}{\xi}$, $\omega_{dB} = \frac{\omega}{\xi}$

and $\xi = \frac{\ln(10)}{10}$

The Moment Generating Function (*MGF*) of the skew normal distribution may be written as [11]:

$$\begin{aligned} M_X(t) &= E\left[e^{tX}\right] \\ &= 2e^{t^2/2} \phi(\beta t), \quad \beta = \frac{\lambda}{\sqrt{1+\lambda^2}} \end{aligned} \tag{13}$$

Thus the first two central moments of $L$ are:

$$\zeta = 2e^\varepsilon e^{\omega^2/2} \phi(\beta \omega) \tag{14}$$

$$\varpi^2 = 2e^{2\varepsilon} e^{\omega^2} (e^{\omega^2} \phi(2\beta\omega) - 2\phi^2(\beta\omega)) \tag{15}$$

## 4. VALIDITY OF LOGNORMAL ASSUMPTION FOR THE SUM OF LOGNORMAL RVS AT DISTRIBUTION TAILS

Several approximation methods for the sum of correlated lognormal RVs is based on the fact that this sum can be approximated, at least as a first order, by another lognormal distribution. On the one hand, Szyszkowicz and Yanikomeroglu [14] have published a limit theorem that states that the distribution of a sum of identically distributed equally and positively correlated lognormal RVs converges, *in distribution*, to a lognormal distribution as N becomes large. This limit theorem is extended in [15] to the sum of correlated lognormal RVs having a particular correlation structure.

On the other hand, some recent results [16, Theorem 1. and 3.] show that the sum of lognormal RVs exhibits a different behaviour at the upper and lower tails even in the case of identically distributed lognormal RVs. Although the lognormal distribution have a symmetric behaviours in both tails, this is not in contradiction with results proven in [14-15] since convergence is proved *in distribution*, i.e., convergence at every point *x* not in the *limit behaviour*.

This explain why some lognormal based methods provide a good accuracy only in the lower tail (e.g. Schwartz and Yeh), where some other methods provide an acceptable accuracy in the upper tail (e.g. Fenton-Wilkinson). This asymmetry of the behaviours of the sum of lognormal RVs at the lower and upper tail motivates us to use the Log Skew Normal distribution as it represents the asymmetric version of lognormal distribution. So, we expect that LSN approximation provide the needed accuracy over the whole region including both tails of the sum of correlated lognormal RVs distribution.

## 5. LOG SKEW NORMAL PARAMETERS DERIVATION

### 5.1. Tails properties of sum of Correlated lognormal RVs

Let $\vec{L}$ be an N-dimensional log-normal vector with parameters $\mu$ and $M$. Let $B = M^{-1}$ the inverse of the covariance matrix.

To study tails behaviour of sum of correlated lognormal RVs, it is convenient to work on lognormal probability scale [13], i.e., under the transformation G:

$$G : F(x) \mapsto \widetilde{F}(x) = F(\phi^{-1}(F(e^x))) \quad (16)$$

We note that under this transformation, the lognormal distribution is mapped onto a linear equation.

We define $B_i$ as row sum of $B$:

$$B_i = \sum_{k=1}^{N} B(i,k), \quad 1 \leq i \leq N \quad (17)$$

Let:

$$\tilde{N} \triangleq Card\{i = 1,..N; \ B_i \neq 0\}$$
$$\tilde{I} \triangleq \{i = 1,..N; \ B_i \neq 0\} \triangleq \{\tilde{k}(1), \tilde{k}(1)...\tilde{k}(\tilde{N})\}$$

We define $\tilde{\mu}$, $\tilde{M}$, $\tilde{B}$ and $\tilde{B}_i$ such that:

$$\tilde{\mu}(i) = \mu(\tilde{k}(i))$$
$$\tilde{M}(i,j) = M(\tilde{k}(i), \tilde{k}(j))$$
$$\tilde{B} = \tilde{M}^{-1}$$
$$\tilde{B}_i = \sum_{k=1}^{\tilde{N}} \tilde{B}(i,k)$$

Since variables $L_i$ are exchangeable, we can assume for the covariance matrix B, with no loss of generality, that $\tilde{I} = \{1, 2, 3...\tilde{N}\}$ with $\tilde{N} \leq N$.

Let $w \in \Re^{\tilde{N}}$ such that:

$$w = \frac{\tilde{B}^{-1}1}{1^{\perp}\tilde{B}^{-1}1} \quad (18)$$

So that, we may write:

$$w_i = \frac{\tilde{A}_i}{\sum_{j=1}^{\tilde{N}} \tilde{A}_j} \quad j = 1,...\tilde{N} \quad (19)$$

Now, we set $\tilde{w} \in \Re^N$ as

$$\tilde{w}_i = \begin{cases} w_i & \text{if } i \leq \tilde{N} \\ 0 & \text{Otherwise} \end{cases} \quad (20)$$

Assuming that for every $i \in \{1,2,3...N\} \setminus \tilde{I}$

$$(e^i - \tilde{w})^\perp B\tilde{w} \neq 0 \quad (21)$$

Where $e^i \in \Re^N$ satisfies $e^i_j = 1$ if $i = j$ and $e^i_j = 0$ otherwise.

In [16, Theorem 3.], Gulisashvili and Tankov proved that the slope of the right tail of the SLN cdf on lognormal probability scale is equal to $1/\underset{i}{\text{Max}}\{\tilde{B}(i,i)\}$ when assumption (20) is valid.

$$\lim_{x \to +\infty} \frac{\delta}{\delta x} \widetilde{F}_{SCLN}(x) = \frac{1}{\underset{i}{\text{Max}}\{\tilde{B}(i,i)\}} \quad (22)$$

Considering the left tail slope, they proved that the slope of the left tail of the SLN cdf on lognormal probability scale is equal to $\sqrt{\sum_{i=1}^{N} \tilde{B}_i}$ [16, Theorem 1.].

$$\lim_{x \to -\infty} \frac{\delta}{\delta x} \widetilde{F}_{SCLN}(x) = \sqrt{\sum_{i=1}^{N} \tilde{B}_i} \quad (23)$$

In general, we can assume that $B_i \neq 0$ for $1 \leq i \leq N$, so that $N = \tilde{N}$ and tails slope can be expressed as:

$$\lim_{x \to +\infty} \frac{\delta}{\delta x} \widetilde{F}_{SCLN}(x) = \frac{1}{\underset{i}{\max}\{B(i,i)\}} \quad (24)$$

$$\lim_{x \to -\infty} \frac{\delta}{\delta x} \widetilde{F}_{SCLN}(x) = \sqrt{\sum_{i=1}^{N} B_i} \quad (25)$$

### 5.2. Tail properties of Skew Log Normal

In [17], it has been showed that the rate of decay of the right tail probability of a skew normal distribution is equal to that of a normal variate, while the left tail probability decays to zero faster. This result has been confirmed in [18]. Based on that, it is easy to show that the rate of decay of the right tail probability of a log skew normal distribution is equal to that of a lognormal variate. Under the transformation G, skew lognormal distribution has a linear asymptote in the upper limit with slope

$$\lim_{x \to +\infty} \frac{\delta}{\delta x} \widetilde{F}_{LSN}(x) = \frac{1}{w} \quad (26)$$

In the lower limit, it has no linear asymptote, but does have a limiting slope

$$\lim_{x \to -\infty} \frac{\delta}{\delta x} \widetilde{F}_{LSN}(x) = \frac{\sqrt{1+\lambda^2}}{w} \quad (27)$$

These results are proved in [8, Appendix A]. Therefore, it will be possible to match the tail slopes of the LSN with those of the sum of correlated lognormal RVs distribution in order to find LSN optimal parameters.

**5.3. Moments and lower tail slope matching**

In order to derive log skew normal optimal parameters, we proceed by matching the two central moments of both distributions. Furthermore, use we lower slope tail match. By simulation, we point out that upper slope tail match is valid only for the sum of high number of lognormal RVs. However we still need it to find an optimal starting guess solution to the generated nonlinear equation. Thus we define $\lambda_{opt}$ as solution the following nonlinear equation:

$$\frac{\sum_{i=1}^{N} e^{2\mu_i} e^{\sigma_i^2} (e^{\sigma_i^2} - 1)}{(\sum_{i=1}^{N} e^{\mu_i} e^{\sigma_i^2/2})^2} = e^{\sqrt{\frac{1+\lambda^2}{\sum_{i=1}^{N} \tilde{B}_i}}} \frac{\phi(2\frac{\lambda}{\sqrt{\sum_{i=1}^{N} \tilde{B}_i}})}{2\phi^2(\frac{\lambda}{\sqrt{\sum_{i=1}^{N} \tilde{B}_i}})} - 1 \quad (28)$$

Such nonlinear equation can be solved using different mathematical utility (e.g. fsolve in Matlab). Using upper slope tail match we derive a starting solution guess $\lambda_0$ to (23) in order to converge rapidly (only few iterations are needed):

$$\lambda_0 = \sqrt{\left[ \max_i \{\tilde{B}(i,i)\}^2 \sum_{i=1}^{N} \tilde{B}_i \right] - 1} \quad (29)$$

Optimal location and scale parameters $\varepsilon_{opt}$, $\omega_{opt}$ are obtained according to $\lambda_{opt}$ as.

$$\begin{cases} \omega_{opt} = \sqrt{\frac{1+\lambda_{opt}^2}{\sum_{i=1}^{N} \overline{B}_i}} \\ \varepsilon_{opt} = \ln(\sum_{i=1}^{N} e^{\mu_i} e^{\sigma_i^2/2}) - \frac{\omega_{opt}^2}{2} - \ln(\phi(\frac{\lambda_{opt}}{\sqrt{\sum_{i=1}^{N} \overline{B}_i}})) \end{cases} \quad (30)$$

## 6. SIMULATION RESULTS

In this section, we propose to validate our approximation method and compare it with other widely used approximation methods. Fig. 1 and Fig. 2 show the results for the cases of the sum of 20 independent lognormal RVs ($\rho = 0$) with mean 0dB and standard deviation 3dB and 6dB. The CDFs are plotted in lognormal probability scale [13]. Simulation results show that the accuracy of our approximation get better as the number of lognormal distributions increase.

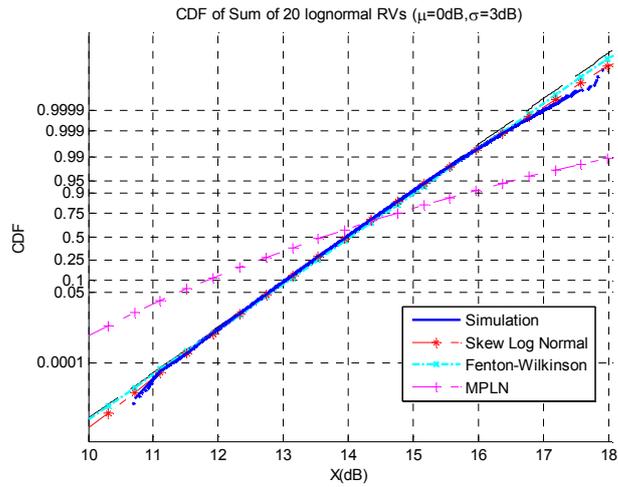

Figure 1. CDF of a sum of 20 i.i.d. lognormal RVs with $\mu$ = 0 dB and $\sigma$ = 3dB.

We can see that LSN approximation offers accuracy over the entire body of the distribution for both cases. In Fig. 3, we consider the sum of 12 independent lognormal RVs having the same standard deviation of 3dB, but with different means. It is clear that LSN approximation catch the entire body of SLN distribution. In this case, both LSN and MPLN provide a tight approximation to SLN distribution. However LSN approximation outperforms MLPN approximation in lower region. Since interferences modelling belongs to this case (i.e. same standard deviation with different means), it is important to point out that log skew normal distribution outperforms other methods in this case.

Fig. 4 shows the case of the sum of 6 independent lognormal RVs having the same mean 0dB but with different standard deviations. We can see that Fenton-Wilkinson approximation method can only fit a part of the entire sum of distribution, while the MPLN offers accuracy on the left part of the SLN distribution. However, it is obvious that LSN method provides a tight approximation to the SLN distribution except a small part of the right tail. It is worthy to note that in all cases, log skew normal distribution provide a very tight approximation to SLN distribution in the region of CDF less than 0.999.

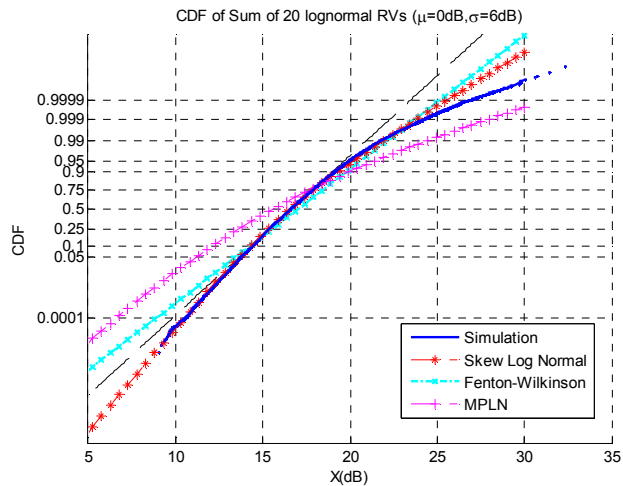

Figure 2. CDF of a sum of 20 i.i.d. lognormal RVs with $\mu$ = 0 dB and $\sigma$ = 6dB.

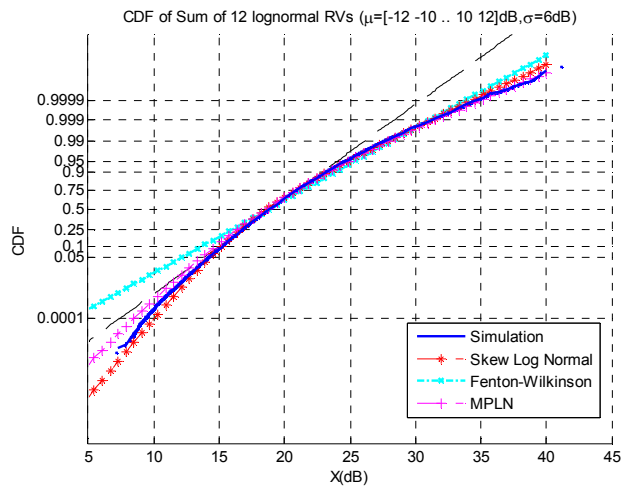

Figure 3. CDF of a sum of 12 lognormal RVs with $\mu$ = [-12 -10 -8 ... 8 10 12] dB and $\sigma$ = 6dB.

The comparison of the Complementary Cumulative Distribution Function (CDF) of the sum of $N$ ($N=2,8,20$) correlated lognormal RVs $P[\Lambda > \lambda]$ of Monte Carlo simulations with LSN approximation and lognormal based methods for low value of standard deviation $\sigma = 3\,dB$ with $\mu = 0\,dB$ and $\rho = 0.7$ are shown in Fig.5. Although these methods are known by its accuracy in the lower range of standard deviation, it obvious that LSN approximation outperforms them. We note that fluctuation at the tail of sum of lognormal RVs distribution is due to Monte Carlo simulation, since we consider $10^7$ samples at every turn. We can see that LSN approximation results are identical to Monte Carlo simulation results.

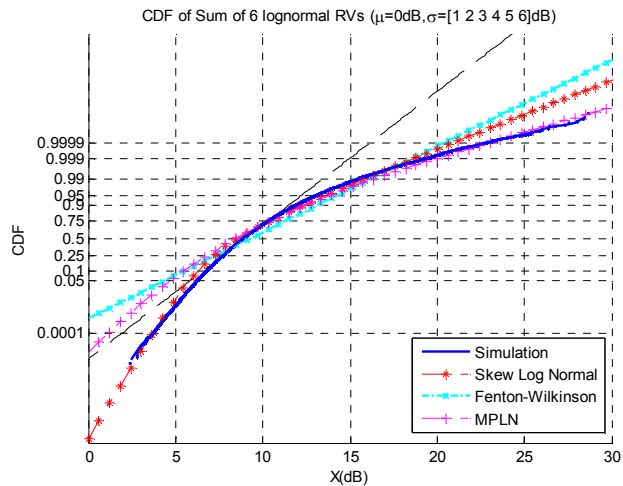

Figure 4. CDF of a sum of 6 lognormal RVs with $\mu$ = 0 dB and $\sigma$ = [1 2 3 4 5 6] dB.

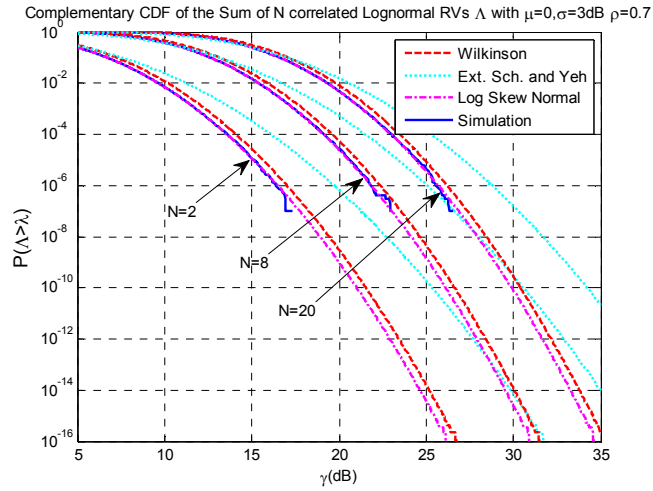

Figure 5. Complementary CDF of the sum of N correlated lognormal RVs with $\mu = 0\,dB$, $\sigma = 3\,dB$ and $\rho = 0.7$

Fig.6 and Fig.7 show the complementary CDF of the sum of *N* correlated lognormal RVs for higher values of standard deviation $\sigma = 6, 9\,dB$ and two values of correlation coefficients $\rho = 0.9, 0.3$. We consider the Log Shifted Gamma approximation for comparison purposes. We consider the Log Shifted Gamma approximation for comparison purposes. We can see that LSN approximation highly outperforms other methods especially at the CDF right tail ($1-CDF < 10^{-2}$). Furthermore, LSN approximation give exact Monte Carlo simulation results even for low range of the complementary CDF of the sum of correlated lognormal RVs ($1-CDF < 10^{-6}$).

To further verify the accuracy of our method in the left tail as well as the right tail of CDF of the sum of correlated lognormal RVs, Fig. 8 and Fig. 9 show respectively the CDF $P[\Lambda < \lambda]$ of the sum of 6 correlated lognormal RVs with $\mu = 0\,dB$ $\rho = 0.7$ and the complementary CDF of the sum of 20 correlated lognormal RVs with $\mu = 0\,dB$ $\rho = 0.3$ for different standard deviation values. It is obvious that LSN approximation provide exact Monte Carlo simulation results at the left tail as well as the right tail. We notice that accuracy does not depend on standard deviation values as LSN performs well for the highest values of standard deviation of the shadowing.

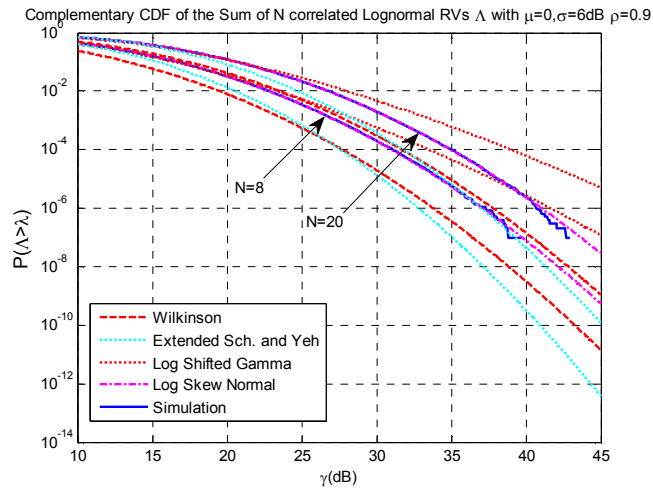

Figure 6. Complementary CDF of the sum of N correlated lognormal RVs with $\mu = 0\,dB$, $\sigma = 6\,dB$ and $\rho = 0.9$

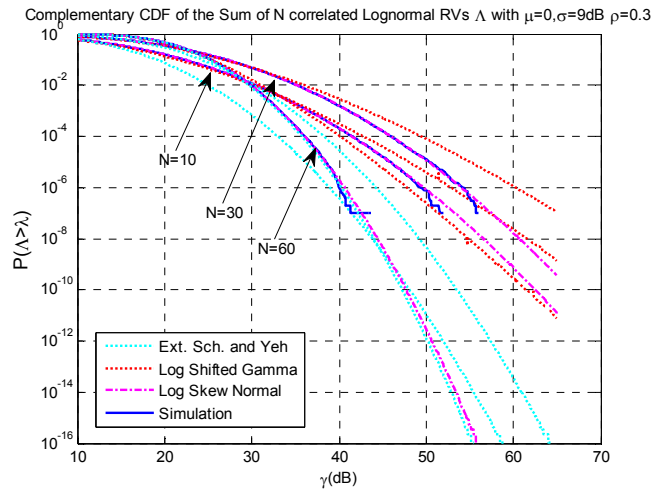

Figure 7. Complementary CDF of the sum of N correlated lognormal RVs with $\mu = 0\,dB$ $\sigma = 9\,dB$ $\rho = 0.3$

To point out the effect of correlation coefficient value on the accuracy of the proposed approximation, we consider the CDF of the sum of 12 correlated lognormal RVs for different combinations of standard deviation and correlation coefficient values with $\mu = 0\,dB$ (Fig. 10). One can see that LSN approximation efficiency does not depend on correlation coefficient or standard deviation values. So that, LSN approximation provides same results as Monte Carlo simulations for all cases.

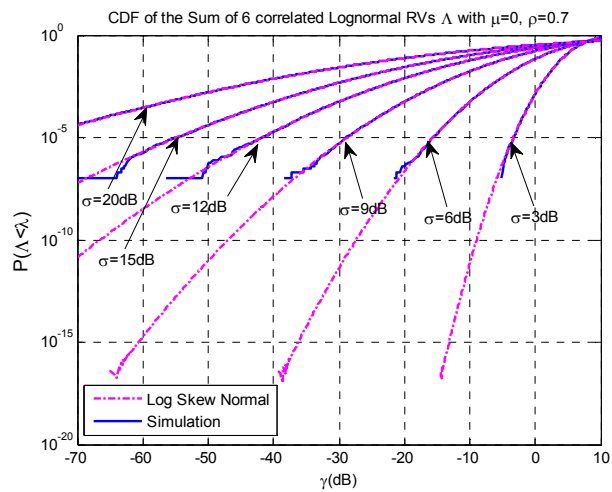

Figure 8. CDF of the sum of 6 correlated lognormal RVs with $\mu = 0\,dB$, $\rho = 0.7$

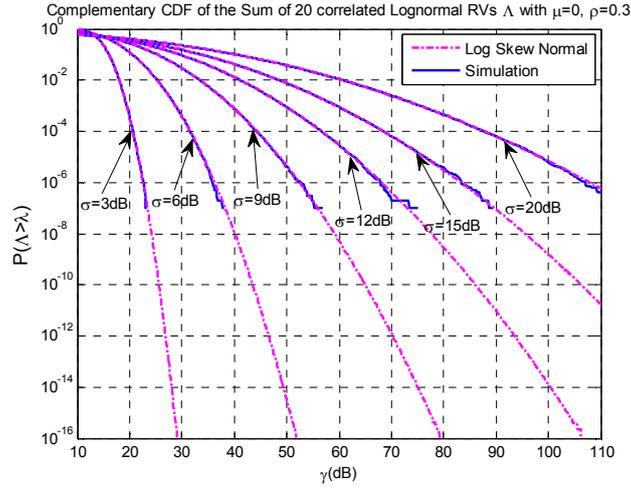

Figure 9. Complementary CDF of the sum of 20 correlated lognormal RVs with $\mu = 0\, dB$, $\rho = 0.3$

## 7. APPLICATION: OUTAGE PROBABILITY IN LOGNORMAL SHADOWING ENVIRONMENT

In this section, we provide an example for outage probability calculation in lognormal shadowing environment based on log skew normal approximation.

We consider a homogeneous hexagonal network made of 18 rings around a central cell. Fig. 11 shows an example of such a network with the main parameters involved in the study: R, the cell range (1 km), Rc, the half-distance between BS. We focus on a mobile station (MS) $u$ and its serving base station (BS), $BS_i$, surrounded by M interfering BS

To evaluate the outage probability, the noise is usually ignored due to simplicity and negligible amount. Only inter-cell interferences are considered. Assuming that all BS have identical transmitting powers, the SINR at the $u$ can be written in the following way:

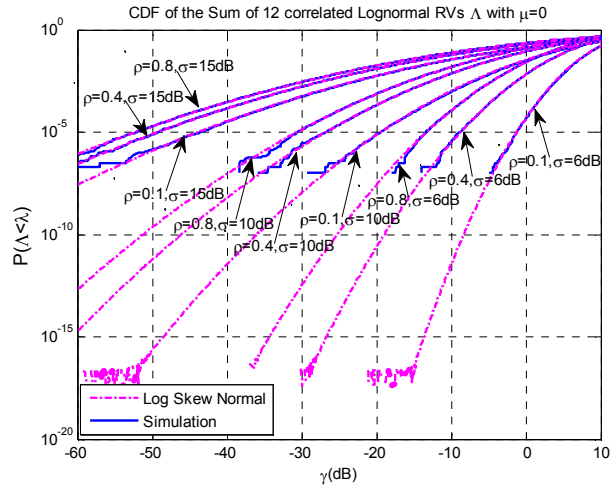

Figure 10. CDF of the sum of 12 correlated lognormal RVs with different correlation coefficients $\rho$, $\mu = 0\, dB$

$$SINR = \frac{S}{I+N} = \frac{P_i K r_{i,u}^{-\eta} \cdot Y_{i,u}}{\sum_{j=0, j\neq i}^{M} P_j K r_{j,u}^{-\eta} \cdot Y_{j,u} + N} = \frac{r_i^{-\eta} \cdot Y_{i,u}}{\sum_{j=0, j\neq i}^{M} r_j^{-\eta} \cdot Y_{j,u}} \quad (31)$$

The path-loss model is characterized by parameters K and $\eta > 2$. The term $P_l K r_l^{-\eta}$ is the mean value of the received power at distance $r_l$ from the transmitter $BS_l$. Shadowing effect is represented by lognormal random variable $Y_{l,u} = 10^{\frac{x_{l,u}}{10}}$ where $x_{l,u}$ is a normal RV, with zero mean and standard deviation σ, typically ranging from 3 to 12 dB.

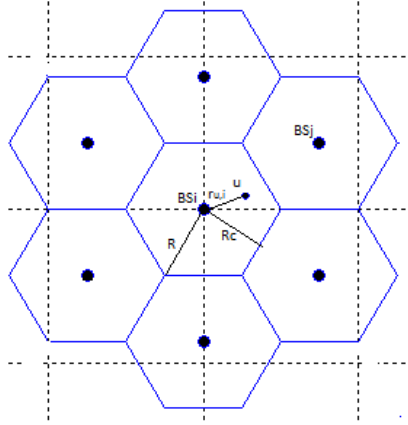

Figure 11. Hexagonal network and main parameters

The outage probability is now defined as the probability for the $\gamma$ SINR to be lower than a threshold value $\delta$:

$$\begin{aligned} P(\gamma < \delta) &= P(\frac{P_{\text{int}}}{P_{\text{ext}}} < \delta) = P(\frac{r_i^{-\eta} \cdot Y_{i,u}}{\sum_{j=0, j\neq i}^{M} r_j^{-\eta} \cdot Y_{j,u}} < \delta) \\ &= P(10\log(\frac{r_i^{-\eta} \cdot Y_{i,u}}{\sum_{j=0, j\neq i}^{M} r_j^{-\eta} \cdot Y_{j,u}}) < \delta_{dB}) \\ &= P(P_{\text{int},dB} - P_{\text{ext},dB} < \xi \delta_{dB}) \end{aligned} \quad (32)$$

Where:

$P_{\text{int,dB}} = \ln(r_i^{-\eta} \cdot Y_{i,u})$ a normal RV with mean $m = \ln(r_i^{-\eta})$ and standard deviation $\xi\sigma$.

$P_{\text{ext,dB}} = \ln(\sum_{j=0, j\neq i}^{M} r_j^{-\eta} \cdot Y_{j,u})$ a skew normal RV with distribution $SN(\lambda, \varepsilon, \omega)$.

It is easy to show that the difference $P_{int,dB} - P_{ext,dB}$ has a skew normal distribution $SN(\lambda_1, \varepsilon_1, \omega_1)$ where:

$$\begin{aligned} \varepsilon_1 &= m - e \\ \omega_1 &= \sqrt{\xi^2 \sigma^2 + \omega^2} \\ \lambda_1 &= \frac{\lambda}{\sqrt{(1+\lambda^2)(\frac{\xi^2 \sigma^2}{\omega^2})+1}} \end{aligned} \qquad (33)$$

Fig. 12 and Fig. 13 show the outage probability at cell edge (r=Rc) and inside the cell (r=Rc/2), resp. for =3dB and 6dB assuming $\eta$=3. Difference between analysis and simulation results is less than few tenths of dB. This accuracy confirms that the LSN approximation, considered in this work, is efficient for interference calculation process.

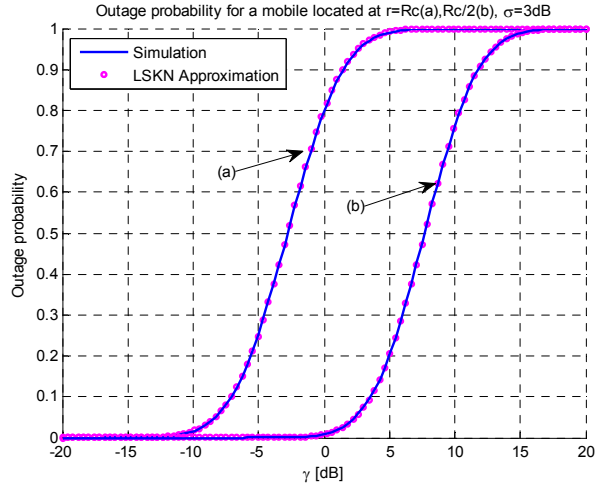

Figure 12. Outage probability for a mobile located at r=Rc (a), r=Rc /2(b), $\sigma$ =3dB, $\eta$=3

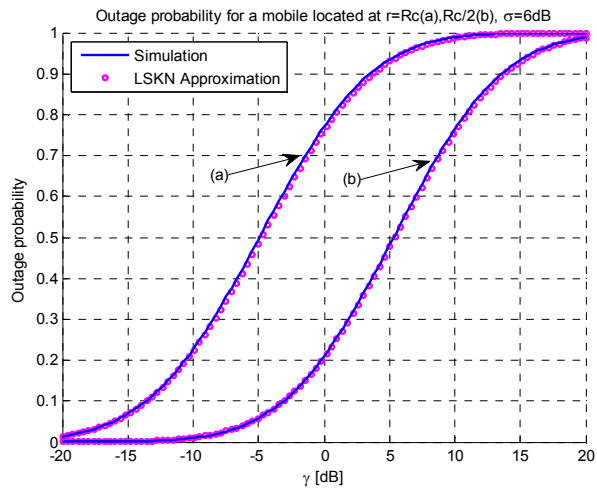

Figure 13. Outage probability for a mobile located at r=Rc (a), r=Rc /2(b), $\sigma$ =6dB, $\eta$=3

## 8. CONCLUSIONS

In this paper, we proposed to use the Log Skew Normal distribution in order to approximate the sum of correlated lognormal RV distribution. Our fitting method uses moment and tails slope matching technique to derive LSN distribution parameters. LSN provides identical results to Monte Carlo simulations results and then outperforms other methods for all cases. Using an example for outage probability calculation in lognormal shadowing environment, we proved that LSN approximation, considered in this work, is efficient for interference calculation process.

**Authors**

**Marwane Ben Hcine** was born in Kébili, Tunisia, on January 02, 1985. He graduated in Telecommunications Engineering, from The Tunisian Polytechnic School (TPS), July 2008. In June 2010, he received the master's degree of research in communication systems of the Higher School of Communication of Tunisia (Sup'Com). Currently he is a Ph.D. student at the Higher School of Communication of Tunisia. His research interests are network design and dimensioning for LTE and beyond Technologies.

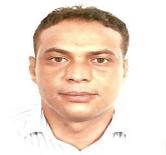

**Pr. Ridha BOUALLEGUE** was born in Tunis, Tunisia. He received the M.S degree in Telecommunications in 1990, the Ph.D. degree in telecommunications in 1994, and the HDR degree in Telecommunications in 2003, all from National School of engineering of Tunis (ENIT), Tunisia. Director and founder of National Engineering School of Sousse in 2005. Director of the School of Technology and Computer Science in 2010. Currently, Prof. Ridha Bouallegue is the director of Innovation of COMmunicant and COoperative Mobiles Laboratory, INNOV'COM Sup'COM, Higher School of Communication. His current research interests include mobile and cooperative communications, Access technique, intelligent signal processing, CDMA, MIMO, OFDM and UWB systems.

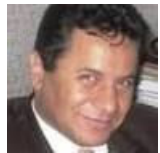